\documentclass[letterpaper,preprintnumbers,eqsecnum,superscriptaddress,aps,nofootinbib]{revtex4}
\usepackage{amssymb}\usepackage[centertags]{amsmath}\usepackage{txfonts}\usepackage{epsfig}\usepackage{bm}
\usepackage{color}\usepackage{graphicx,graphics}\usepackage{multirow}\usepackage{float}\usepackage{ulem}
\usepackage[dvipdfm,pdfstartview=FitH]{hyperref}\usepackage{setspace}\usepackage{slashed}\usepackage{booktabs}
\hypersetup{colorlinks=true, citecolor=blue, linkcolor=blue,filecolor=black,urlcolor=blue}
\allowdisplaybreaks[2]

\begin{document}

\title{Azimuthal asymmetries from $\theta$ Vacuum}

\author{Weihua Yang}


\affiliation{Department of Modern Physics, University of Science and Technology of China, Hefei, Anhui 230026, China}

\begin{abstract}
   In this note, we calculate azimuthal asymmetries in terms of fragmentation functions in semi-inclusive electron positron annihilation process. These fragmentation functions are divided into two parts according to the parity even or parity odd properties. We present eight kinds of azimuthal asymmetries and we categorize them for four categories.
   First, asymmetries comes from both $\mathcal{P}$ odd FFs and $\mathcal{P}$ even FFs which have same signs (e.g., $H^\perp_1\bar H^\perp_1+\mathcal{H}^\perp_1\bar{\mathcal{H}}^\perp_1$). Second, asymmetries comes from two interference terms between the $\mathcal{P}$ odd FF and $\mathcal{P}$ even FF which have opposite signs (e.g., $H^\perp_1\mathcal{H}^\perp_1-\mathcal{H}^\perp_1\bar H^\perp_1$).  Third, asymmetries comes from both $\mathcal{P}$ odd FFs and $\mathcal{P}$ even FFs which have opposite signs (e.g., $\mathcal{G}^\perp_{1T}\bar{\mathcal{D}}_1-D^\perp_{1T}\bar D_1$). Fourth, asymmetries comes from two interference terms between the $\mathcal{P}$ odd FF and $\mathcal{P}$ even FF which have same signs ($H_{1T}\bar {\mathcal{H}}^\perp_1+\mathcal{H}_{1T}\bar H_1^\perp$ ).
   The $\mathcal{P}$-odd fragmentation functions, e.g., $\mathcal{H}_1^\perp$, are induced by $local$ $\mathcal{P}$- and/or $\mathcal{CP}$-odd effects which are connected to the tunneling ($\theta$ vacuum) events. Due to the $\mathcal{P}$-odd effect, azimuthal asymmetries induced by the effect vanish if all events are summed over but survive on the event-by-event for the interference terms with opposite signs. However, interference terms with same signs survive and can be used to measure/study the $\mathcal{P}$-odd fragmentation functions. Finally, we also calculate hadron polarizations in terms of both the $\mathcal{P}$-even and $\mathcal{P}$-odd fragmentation functions.
\end{abstract}


\maketitle

\section{Introduction}\label{Sec-introduction}

Parton distribution functions (PDFs) and fragmentation functions (FFs) are two important quantities in describing the high energy reactions. Both of them are long distance non-perturbative quantities which cannot be calculated with perturbative theory, they are mainly calculated by phenomenological models and parametrized by experiment data~\cite{Jakob:1997wg,Avakian:2010br,Bacchetta:2001di,Amrath:2005gv,Metz:2002iz,Bacchetta:2007wc,Davidson:2001cc, Ito:2009zc,Matevosyan:2010hh,Matevosyan:2011vj,Matevosyan:2012ga,Dulat:2015mca,Albino:2008fy,canshuhua,deFlorian:1997zj}. When three-dimensional, i.e., the transverse momentum dependent (TMD) PDFs and FFs are considered, the sensitive quantities studied in experiments are often different azimuthal asymmetries.
In the quantum field theoretical formulation, PDFs and/or FFs are defined by the quark-quark correlators (correlation functions)  which are defined as $4\times 4$ matrices in Dirac space depending on the hadron state. As a result, the correlators can be decomposed by the Dirac matrices, i.e., $\Gamma=\{I, i\gamma^5, \gamma^\mu, \gamma^\mu\gamma^5, i\sigma^{\mu\nu}\gamma^5\}$. The corresponding coefficients can be further decomposed by the Lorentz covariants and scalar functions and these scalar functions are called PDFs and FFs.

Generally, the quark-quark correlator satisfies two constraints, Hermiticity and parity conservation. Hermiticity ensures that all the scalar functions are real while parity conservation strictly limits the numbers of scalar functions. Even though parity violating effects are not expected in perturbative quantum chromodynamics (QCD), the non-trivial $\theta$ vacuum may lead to the constraint breakdown~\cite{tHooft:1976rip,tHooft:1976snw,Jackiw:1976pf,Callan:1976je}. The $\theta$ term enters in the lagrangian as
\begin{align}
  \mathcal{L}_\theta=\frac{\theta}{16\pi^2}\mathrm{Tr}\big[F_{\mu\nu}\tilde F^{\mu\nu}\big] \label{f:Lagrangian}
\end{align}
where $F_{\mu\nu}$ is the gluon field strength tensor and $\tilde F^{\mu\nu}=\varepsilon^{\mu\nu\rho\sigma}F_{\rho\sigma}/2$. We see that the $\theta$ term in lagrangian violates the  parity ($\mathcal{P}$) and/or charge-parity ($\mathcal{CP}$) symmetries.
However the measurements of electric dipole moment of neutron indicate that the parity ($\mathcal{P}$) violation is $local$~\cite{Baker:2006ts}. Even though QCD is $local$ $\mathcal{P}$ violated, Kharzeev, Pisarski and Tytgat have shown it can be directly observed~\cite{Kharzeev:1998kz}. In heavy ion collisions, it also leads to the famous chiral magnetic effect~\cite{Kharzeev:2004ey,Kharzeev:2007jp,Fukushima:2008xe}. Furthermore, Efremov, Kharzeev and Kang also discussed the $local$ $\mathcal{P}$ odd effect in fragmentation process. In this case, the $\mathcal{P}$ odd FFs emerge and they argued that the effect can be detected in experiments via physical observables, e.g., handedness correlation and azimuthal asymmetries~\cite{Efremov:1995ff,Kang:2010qx}.

Electron positron annihilation process is regard as the most cleanest place to study the FFs, because in this process there are no hadronic effects in initial states. As we mentioned before the sensitive quantities studied in experiments are often different azimuthal asymmetries when TMD FFs are considered. In ref.~\cite{Kang:2010qx}, Kang and Kharzeev derived the most general form of the FF for quarks fragmenting into spin-0 hadrons (pseudoscalars) at leading twist. Without the parity constraint they found two more $\mathcal{P}$ odd FFs which are related to two azimuthal asymmetries, i.e., $\cos2\phi$ and $\sin2\phi$. They presented a estimation of the magnitude of the $\sin2\phi$ asymmetry, they found $I(\bar\theta,z_1,z_2)\sim 1.5\%$. This is a relative large asymmetry and can be detected.
In this note we extend the calculation to spin-1/2 dihadron production. We calculate the azimuthal asymmetries for unpolarized, single-spin asymmetries and double spin asymmetries, respectively. We also calculate the hadron polarizations.

This note is organized as follows. In Sect.~\ref{Sec-kinematics} we present the kinematics of the semi-inclusive electron positron annihilation (SIA) precess and the conventions and notations used in this note. In Sect.~\ref{Sec-correlations} we present the decompositions of the correlation functions as well as the complete hadronic tensor. Our calculation results are shown in Sect.~\ref{Sec-crosssection}. And a brief summary will be given in Sect.~\ref{Sec-summary}.

\section{Kinematics}\label{Sec-kinematics}

\begin{figure}[t]
  \includegraphics[width=6cm]{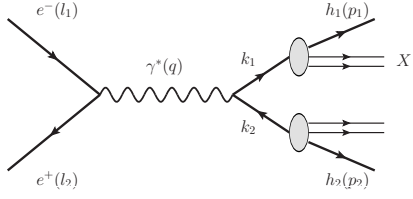}\\
  \caption{Semi-inclusive electron positron annihilation process. The momenta of particles are shown in the parenthesises in the figure.}\label{elepos}
\end{figure}

In this section we present the kinematics of the SIA process, see Fig.~\ref{elepos}. Here SIA process denotes that two hadron are produced in two back-to back-jets in electron positron annihilation preocess, respectively. In this note we only consider the electromagnetic interaction for distinguishing the $\mathcal{P}$-odd effect in QCD from weak interaction theory. The differential cross section of SIA process can be written as contraction of the leptonic tensor and hadronic tensor, i.e.,
\begin{align}
  \frac{2E_1E_2d\sigma}{d^3\vec p_1d^3\vec p_2}=\frac{N_c\alpha^2_{em}e_q^2}{sQ^4}L_{\mu\nu}(l_1,l_2)W^{\mu\nu}(p_1, p_2),\label{f:cross-pp}
\end{align}
where $N_c$ denotes the color factor, $\alpha_{em}$ is the fine structure constant and $e_q$ is the electric charge of quark $q$.
$s=Q^2=q^2=(l_1+l_2)^2$. The leptonic tensor is given by
\begin{align}
  L_{\mu\nu}(l_1,l_2)
  &=\frac{1}{4}\mathrm{Tr}\left[\slashed l_1 \gamma_\mu \slashed l_2 \gamma_\nu\right]\nonumber\\
  &=l_{1\mu} l_{2\nu}+l_{1\nu} l_{2\mu}-g_{\mu\nu}\left(l_1\cdot l_2\right).\label{f:leptonictensor}
\end{align}
The operator definition of hadronic tensor is given by
\begin{align}
  &W^{\mu\nu}(p_1,p_2)=\Delta\sum_X\langle 0|J^\mu(0)|p_1, p_2;X\rangle\langle p_1,p_2;X|J^\nu(0)|0\rangle \nonumber\\
  &=\int d^2ye^{-iqy}\sum_X\langle 0|J^\mu(0)|p_1, p_2;X\rangle \langle p_1, p_2;X|J^\nu(y)|0\rangle ,\label{f:hadronictensor}
\end{align}
where $\Delta=\delta^4(q-p_1-p_2-X)$. Besides the Lorentz covariance, the hadronic tensor $W_{\mu\nu}$ satisfies the general constraints imposed by hermiticity, current conservation, and parity conservation in the electromagnetic process, i.e.,
\begin{align}
  W^{*\mu\nu}(p_1,S_1; p_2,S_2)&=W^{\nu\mu}(p_1,S_1; p_2,S_2), \label{f:WHermiticity} \\
  q_\mu W^{\mu\nu}(p_1,S_1; p_2,S_2)&=q_\nu W^{\mu\nu}(p_1,S_1; p_2,S_2)=0, \label{f:WCurrent} \\
  W^{\mu\nu}(p_1,S_1; p_2,S_2)&= W_{\mu\nu}(p^\mathcal{P}_1,S^\mathcal{P}_1; p^\mathcal{P}_2,S^\mathcal{P}_2), \label{f:WParity}
\end{align}
where a vector with the superscript $\mathcal{P}$ denotes the result after space reflection such as $p_\mu^\mathcal{P}=p^\mu$.
From Eq.~(\ref{f:hadronictensor}) we know that the hadronic tensor can not be calculated perturbatively because it contains the nonperturbative hadronization process. However, to obtain the cross section, we can decompose the hadronic tensor with the basic Lorentz tensors as shown in ref.~\cite{Pitonyak:2013dsu,Chen:2016moq}. Here we will not repeat the decompositions.

\begin{figure}[t]
  \includegraphics[width=6cm]{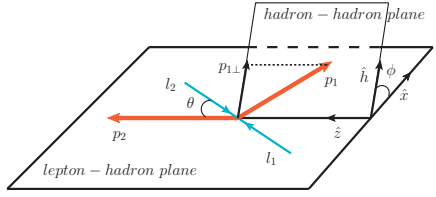}\\
  \caption{Semi-inclusive electron positron annihilation process in the lepton center of mass frame.}\label{Frame}
\end{figure}

To calculate the SIA process, we choose the lepton center-of-mass frame where $p_2$ lies in the positive $z$ direction, see Fig.~\ref{Frame}.  In this frame we use the following conventions and notation given in ref.~\cite{Boer:1997mf}. We define
\begin{align}
\hat t^\mu& \equiv  \frac{q^\mu}{Q},  \label{f:t} \\
\hat z^\mu& \equiv  \frac{Q}{p_2\cdot q} p^\mu_{2q} =2\frac{p_2^\mu}{z_2 Q} - \frac{q^\mu}{Q}, \label{f:z}
\end{align}
where we have defined $p_{q}^\mu=p^\mu-\frac{p\cdot q}{q^2}q^\mu$ and the higher twist terms ($1/Q^2$) is omitted. There are two projectors which can be defined in terms of $\hat t$ and $\hat z$, they are
\begin{align}
& g_{\perp}^{\mu\nu}\equiv  g^{\mu\nu} -\hat t^\mu \hat t^\nu +\hat z^\mu \hat z^\nu, \label{f:gperp}\\
& \varepsilon_\perp^{\mu\nu} \equiv-\epsilon^{\mu\nu\rho\sigma} \hat t_\rho \hat z_\sigma =\frac{1}{p_2\cdot q}\epsilon^{\mu \nu \rho\sigma} p_{2\,\rho}q_\sigma. \label{f:epsilonperp}
\end{align}
Then we have $p_{1\perp}^\mu=g_{\perp}^{\mu\nu}p_{1\nu}$. For convenience, we also define $\hat h^\mu =p_{1\perp}^\mu/|p_{1\perp}|$. FFs are often define with respect to the transverse momentum of the quark, so it would be convenient to define lightlike directions in terms of the momenta of produced hadrons, $p_1, p_2$. Then we have
\begin{align}
& p_1^\mu \equiv \frac{z_1 Q}{\sqrt{2}}n^\mu+ \frac{M_1^2}{z_1Q\sqrt{2}}\bar n^\mu,\\
& p_2^\mu \equiv \frac{M_2^2}{z_2Q\sqrt{2}}n^\mu +\frac{z_2Q}{\sqrt{2}}\bar n^\mu,\\
& q^\mu \equiv \frac{Q}{\sqrt{2}}n^\mu+ \frac{Q}{\sqrt{2}}\bar n^\mu + q_T^\mu,
\end{align}
where $q_T^2 \equiv -Q_T^2$, $M_1, M_2$ are masses of hadron 1 and hadron 2, respectively. Here  we have assumed that $Q_T^2\ll Q^2$. $\bar n$ and $n$ are lightlike unit vectors which satisfy $\bar n\cdot n=1$, $\bar n^2=n^2=0$. In Eq.~(\ref{f:z}), $p_2$ is used to define $\hat z$, then we have $p_{1\perp}^\mu=-z_1q_T^\mu$. We may also define the projectors in terms of $\bar n, n$,
\begin{align}
& g^{\mu\nu}_T \equiv  g^{\mu\nu}- \bar n^{\mu} n^{\nu}-\bar n^{\nu} n^{\mu}, \label{f:gT}\\
& \varepsilon^{\mu\nu}_T \equiv \varepsilon^{\mu\nu\rho\sigma} \bar n_{\rho}n_{\sigma}. \label{f:epsilonT}
\end{align}
We see that the $perpendicular$ vectors and tensors (labeled by $\perp$) are different from the $transverse$ vectors and tensors (labeled by $T$). They can be connected by the following equations, e.g.,
\begin{align}
  & g_T^{\mu\nu} = g_\perp^{\mu\rho}g^{\nu}_{T\rho} -\frac{Q_T}{Q}(\hat z+\hat t)\hat h^\mu, \label{f:gtperp}\\
  & k_\perp^\mu= k_T^\mu-\frac{q_T\cdot k_T}{Q}(\hat z+\hat t), \label{f:kperpt}
\end{align}
where $q_T\cdot k_T=-\mathbf{q}_T\cdot \mathbf{k}_T$. Using Eq.~(\ref{f:kperpt}), we can transform the vectors in transverse basis into perpendicular basis to calculate the hadronic tensor and/or cross section. Calculations will be given in the followings.

On the perpendicular basis the lepton tensor can be parameterized as
\begin{align}
  L_{\mu\nu}=\frac{Q^2}{2}&\Big[-g_\perp^{\mu\nu} + 4y(1-y) \hat z^\mu \hat z^\nu- 4y(1-y)\hat x^\mu\hat x^\nu \nonumber\\
  &-2(1-2y)\sqrt{y(1-y)}\hat z^{\{\mu}\hat x^{\nu\}}\Big] \nonumber\\
  =Q^2&\Big[-A(y)g_\perp^{\mu\nu} +2B(y)\hat z^\mu \hat z^\nu-2B(y)\big(\hat x^\mu\hat x^\nu +\frac{1}{2}g_\perp^{\mu\nu} \big)\nonumber\\
  &- C(y)D(y)\hat z^{\{\mu}\hat x^{\nu\}}\Big], \label{f:Lmunu}
\end{align}
where $\hat x^\mu=l_\perp^\mu/D(y)Q$, and
\begin{align}
A(y) &= \frac{1}{2}-y+y^2, \label{f:Ay}\\
B(y) &= y\,(1-y), \label{f:By}\\
C(y) &= 1-2y, \label{f:Cy}\\
D(y) &= \sqrt{y\,(1-y)}.\label{f:Dy}
\end{align}
Eq.~(\ref{f:Lmunu}) can be used to calculated the cross section by contracting with the hadronic tensor obtained in the following.

\section{Correlation functions and hadronic tensor}\label{Sec-correlations}

\begin{figure}[ht]
  \centering
  \includegraphics[width=4cm]{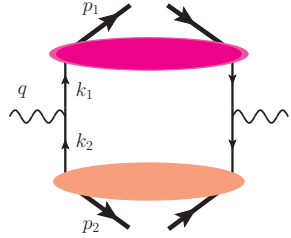}\\
  \caption{The hadronic tensor in parton model.}\label{HTensor}
\end{figure}

Because the hadronic tensor contains nonperturbative process, it can not be calculated by perturbative theory. In this case we should resort to the parton model~\cite{Feynman:1969ej,Feynman:1973xc,Bjorken:1969ja}.
In QCD parton model, see Fig.~\ref{HTensor}, the hadronic tensor can be rewritten as
\begin{align}
  W^{\mu\nu}(p_1, p_2)=&\int\frac{d^2 k_{1T}}{(2\pi)^2}\frac{d^2 k_{2T}}{(2\pi)^2}\delta^2( \vec q_T- \vec k_{1T}- \vec k_{2T})\nonumber\\
  &\mathrm{Tr}\left[\bar \Delta(p_2,k_{2T})\gamma^\mu \Delta(p_1,k_{1T})\gamma^\mu\right], \label{f:hadronTrace}
\end{align}
where $\Delta(z_1,k_{1T})$ and $\bar \Delta(z_2,k_{2T})$ are correlation functions and the operator definitions are given by
\begin{align}
  \Delta(p_1,k_{1T})&=\frac{1}{2\pi}\int d\xi_1^-d^2 \xi_{1T}e^{-ik_1\xi_1}\nonumber\\
  &\sum_X\langle 0|\psi(0)|p_1,X\rangle\langle p_1,X|\bar\psi(\xi_1)|0\rangle|_{\xi_1^+=0},\label{f:Xi1}\\
  \bar\Delta(p_2,k_{2T})&=\frac{1}{2\pi}\int d\xi_2^-d^2 \xi_{2T}e^{-ik_2\xi_2}\nonumber\\
  &\sum_X\langle 0|\bar\psi(0)|p_2,X\rangle\langle p_2,X|\psi(\xi_2)|0\rangle|_{\xi_2^-=0}.\label{f:Xi2}
\end{align}
Equations~(\ref{f:Xi1})-(\ref{f:Xi2}) are not physical correlation functions, because they are not gauge invariant since the spinor field are not located at the same point. To obtain the gauge invariant correlation functions, gauge links should be taken into consideration. Gauge link (also called Wilson line) is defined as
\begin{align}
  \mathcal{L}(0,\infty)=\mathcal{P}exp\big[ig\int^\infty_0 dz n A(y+zn)\big]. \label{f:gaugelink}
\end{align}

Armed with the gauge links, the gauge invariant correlation functions are given by,
\begin{align}
  \hat\Xi(p_1,k_{1T})&=\frac{1}{2\pi}\int d\xi_1^-d^2 \xi_{1T}e^{-ik_1\xi_1}\sum_X\langle 0|\psi(0)\mathcal{L}(0,\infty)\nonumber\\
  &\times|p_1,X\rangle\langle p_1,X|\bar\psi(\xi_1)\mathcal{L}^\dag(\xi,\infty)|0\rangle|_{\xi_1^+=0},\label{f:Xi1gauge}\\
  \bar{\hat\Xi}(p_2,k_{2T})&=\frac{1}{2\pi}\int d\xi_2^-d^2 \xi_{2T}e^{-ik_2\xi_2}\sum_X\langle 0|\bar\psi(0)\mathcal{L}^\dag(0,\infty)\nonumber\\
  &\times|p_2,X\rangle\langle p_2,X|\psi(\xi_2)\mathcal{L}(\xi,\infty)|0\rangle|_{\xi_2^-=0}.\label{f:Xi2gauge}
\end{align}
As we have mentioned in Sect.~\ref{Sec-introduction}, these correlation functions given by Eqs.~(\ref{f:Xi1gauge})-(\ref{f:Xi2gauge}) satisfy the following constraints
imposed by Hermiticity and parity conservation, i.e.,
\begin{align}
  \hat\Xi^\dag(p,k_{T})&=\gamma^0\hat\Xi^\dag(p,k_{T})\gamma^0, \label{f:Hermiticity}\\
  \hat\Xi(p,k_{T})&=\gamma^0\hat\Xi^\dag(p^\mathcal{P},k^\mathcal{P}_{T})\gamma^0. \label{f:Hermiticity}
\end{align}

Correlation functions can not be calculated with perturbative theory because they contain the hadronization information. However,  they are $4\times 4$ matrices in Dirac space and can be decomposed in terms of the $\Gamma$ matrices, i.e., $\Gamma = \{I, i\gamma^5, \gamma^\rho, \gamma^\rho\gamma^5, i\sigma^{\rho\sigma}\gamma^5$\}. The decomposition can be written explicitly as
\begin{align}
  \hat\Xi=I\Xi+i\gamma^5\tilde\Xi +\gamma^\alpha \Xi_\alpha +\gamma^\alpha\gamma^5 \tilde\Xi_\alpha+i\sigma^{\alpha\beta}\gamma^5\Xi_{\alpha\beta}. \label{f:XiD}
\end{align}
Furthermore, with the parity constraint the coefficients in Eq.~(\ref{f:XiD}) can be rewritten as
\begin{align}
  &z\Xi_\alpha=\bar n_\alpha\bigg[D_1+\frac{\varepsilon_{T kS}}{M}D^\perp_{1T}\bigg],\label{f:zxi}\\
  &z\tilde\Xi_\alpha=\bar n_\alpha\bigg[\lambda G_{1L}+\frac{k_T \cdot S_T}{M}G^\perp_{1T}\bigg],\label{f:zxi5}\\
  &z\Xi_{\rho\alpha}=\bar n_\alpha\bigg[\frac{\varepsilon_{T k\alpha}}{M}H^\perp_{1} +S_{T\alpha}H_{1T}+\frac{k_{T\alpha}}{M}H^\perp_{1S}\bigg],\label{f:zxiodd}
\end{align}
at leading twist. We see that $\Xi$ and $\tilde\Xi$ do not have leading twist contributions.
Here we have defined the shorthanded notation $H^\perp_{1S}=\lambda H^\perp_{1L}+\frac{k_T \cdot S_T}{M}H^\perp_{1T}$.
If the parity constraint is released, in other words we consider the $\mathcal{P}$ odd effects caused by the non-trivial structure QCD $\theta$ vacuum, counterparts of the coefficients in the decompositions can be obtained, they are
\begin{align}
  &z\Xi^P_\alpha=\bar n_\alpha\bigg[i\gamma^5\mathcal{D}_1+\frac{k_T \cdot S_T}{M}\mathcal{D}^\perp_{1T}\bigg],\label{f:pzxi}\\
  &z\tilde\Xi^P_\alpha=\bar n_\alpha\bigg[-i\gamma^5\lambda \mathcal{G}_{1L}+\frac{\varepsilon_{T kS}}{M}\mathcal{G}^\perp_{1T}\bigg],\label{f:pzxi5}\\
  &z\Xi^P_{\rho\alpha}=\bar n_\alpha\bigg[\frac{k_{T\alpha}}{M}\mathcal{H}^\perp_{1} +\varepsilon_{T S\alpha}\mathcal{H}_{1T} +\frac{\varepsilon_{T k\alpha}}{M} \mathcal{H}^\perp_{1S}\bigg],\label{f:pzxiodd}
\end{align}
where the superscript $P$ denote the parity violating and $\mathcal{H}^\perp_{1S}=\lambda \mathcal{H}^\perp_{1L}+\frac{k_T \cdot S_T}{M}\mathcal{H}^\perp_{1T}$.  Here we call $D, G, H, \mathcal{D}, \mathcal{G}$ and $ \mathcal{H}$ fragmentation functions in Eqs.~(\ref{f:zxi})-(\ref{f:pzxiodd}). We also use $D, G, H$ to denote the parity conserved FFs and use $\mathcal{D}, \mathcal{G}, \mathcal{H}$ to denote the parity violated FFs. The $\mathcal{P}$-odd and even FFs can be written in a unified form, in this case the correlation function  can be written as
\begin{align}
  z\Xi &=\slashed{\bar n}\bigg[D_1+\lambda \mathcal{G}_{1L}+\frac{\varepsilon_{T kS}}{M}D^\perp_{1T}+\frac{k_T \cdot S_T}{M}\mathcal{D}^\perp_{1T}\bigg]\nonumber\\
  &+\slashed{\bar n}\gamma^5\bigg[\mathcal{D}_1+\lambda G_{1L}+\frac{\varepsilon_{T kS}}{M}\mathcal{G}^\perp_{1T}+\frac{k_T \cdot S_T}{M}G^\perp_{1T}\bigg] \nonumber\\
  &+i\sigma^{\rho\alpha}\gamma^5\bar n_\rho
  \bigg[\frac{\varepsilon_{T k\alpha}}{M}\big(H^\perp_{1}+\mathcal{H}^\perp_{1S}\big) +S_{T\alpha}H_{1T}\nonumber\\
  &~~~~~~~~+\frac{k_{T\alpha}}{M}\big(\mathcal{H}^\perp_{1}+H^\perp_{1S}\big)+\varepsilon_{T S\alpha}\mathcal{H}_{1T} \bigg].\label{f:decompostion}
\end{align}
We have 16 fragmentation functions, 8 parity conserved and 8 parity violated. For every parity conserved one there is a parity violated one corresponding to it, see Table.~\ref{tableFF}. Here we should know that $D_{1T}^\perp$ and $\mathcal{G}_{1T}^\perp$ have the same forms while $G_{1T}^\perp$ and $\mathcal{D}_{1T}^\perp$ have the same forms.

\begin{table}[ht]
 \renewcommand\arraystretch{1.8}
\begin{tabular}{|c||c|c|c|c|c|c|c|c|}
  \hline
  $ \mathcal{P}$-even & $D_1$ & $D_{1T}^\perp$ & $G_{1L}$ & $G_{1T}^\perp$  & $H_{1}^\perp$ & $H_{1T}$ & $H_{1L}^\perp$ & $H_{1T}^\perp$ \\ \hline
  $ \mathcal{P}$-odd  & $\mathcal{D}_1$ & $\mathcal{D}_{1T}^\perp$ & $\mathcal{G}_{1L}$ & ~$\mathcal{G}_{1T}^\perp$ ~ & ~$\mathcal{H}_{1}^\perp$ ~&~ $\mathcal{H}_{1T}$ ~& ~ $\mathcal{H}_{1L}^\perp$ ~& ~$\mathcal{H}_{1T}^\perp$~  \\ \hline
\end{tabular} \caption{$\mathcal{P}$-even and $\mathcal{P}$-odd FFs.} \label{tableFF}
\end{table}


Since we have all the decompositions of the correlation functions, we can calculate the hadronic tensor. To obtain the complete hadronic tensor, we substitute FFs decomposed in Eq.~(\ref{f:decompostion}) into Eq.~(\ref{f:hadronTrace}) (where correlation functions should be changed to the gauge invariant ones) and carry out the calculations. Thus the complete hadronic tensor in SIA process at leading twist is given by
\begin{widetext}
\begin{align}
  &W^{\mu\nu}=\frac{4}{ z_1z_2}\int \frac{d^2 k_{1T}}{(2\pi)^2}\frac{d^2 k_{2T}}{(2\pi)^2}\delta^2( \vec q_T- \vec k_{1T}- \vec k_{2T}) \nonumber\\
  & \times \Bigg\{d_{T}^{\mu\nu}\bigg\{\Big[D_1+\lambda_1 \mathcal{G}_{1L}+\frac{\varepsilon_{T k_1S_1}}{M}D^\perp_{1T}+\frac{k_{1T} \cdot S_{1T}}{M}\mathcal{D}^\perp_{1T}\Big]\Big[\bar D_1+\lambda_2 \bar{\mathcal{G}}_{1L}+\frac{\varepsilon_{T k_2S_2}}{M}\bar D^\perp_{1T}+\frac{k_{2T} \cdot S_{2T}}{M}\bar{\mathcal{D}}^\perp_{1T}\Big]\nonumber\\
  &\hspace{8mm}+\Big[\mathcal{D}_1+\lambda_1 G_{1L}+\frac{\varepsilon_{T k_1S_1}}{M}\mathcal{G}^\perp_{1T}+\frac{k_{1T} \cdot S_{1T}}{M}G^\perp_{1T}\Big]\Big[\bar{\mathcal{D}}_1+\lambda_2 \bar G_{1L}+\frac{\varepsilon_{T k_2S_2}}{M}\bar{\mathcal{G}}^\perp_{1T}+\frac{k_{2T} \cdot S_{2T}}{M}\bar G^\perp_{1T}\Big]\bigg\}\nonumber\\
  &-\frac{d_{T}^{\mu\nu}k_{1T}\cdot k_{2T}+k_{1T}^{\{\mu}k_{2T}^{\nu\}}}{M_1M_2} \big(H^\perp_{1}+\mathcal{H}^\perp_{1S}\big) \big(\bar H^\perp_{1}+\bar{\mathcal{H}}^\perp_{1S}\big) -\frac{d_{T}^{\mu\nu}\varepsilon_{T k_1k_2}+\varepsilon_{T k_1}^{~~~\{\mu}k_{2T}^{\nu\}}}{M_1M_2} \big(H^\perp_{1}+\mathcal{H}^\perp_{1S}\big)\big(\bar {\mathcal{H}}^\perp_{1}+\bar{H}^\perp_{1S}\big) \nonumber\\
  &-\frac{d_{T}^{\mu\nu}k_{1T}\cdot k_{2T}+k_{1T}^{\{\mu}k_{2T}^{\nu\}}}{M_1M_2} \big(\mathcal{H}^\perp_{1}+H^\perp_{1S}\big)\big(\bar{\mathcal{H}}^\perp_{1}+\bar H^\perp_{1S}\big) -\frac{d_{T}^{\mu\nu}\varepsilon_{T k_2k_1}+\varepsilon_{T k_2}^{~~~\{\mu}k_{1T}^{\nu\}}}{M_1M_2} \big(\mathcal{H}^\perp_{1}+H^\perp_{1S}\big)\big(\bar {H}^\perp_{1}+\bar{\mathcal{H}}^\perp_{1S}\big) \nonumber\\
  &-\frac{d_{T}^{\mu\nu}\varepsilon_{T k_1S_2}+\varepsilon_{T k_1}^{~~~\{\mu}S_{2T}^{\nu\}}}{M_1}\big(H^\perp_{1}+\mathcal{H}^\perp_{1S}\big)\bar{H}_{1T} -\frac{d_{T}^{\mu\nu}k_{1T}\cdot S_{2T}+k_{1T}^{\{\mu}S_{2T}^{\nu\}}}{M_1} \big(H^\perp_{1}+\mathcal{H}^\perp_{1S}\big)\bar{\mathcal{H}}_{1T} \nonumber\\
  &- \frac{d_{T}^{\mu\nu}\varepsilon_{T S_2k_1}+\varepsilon_{T S_2}^{~~~\{\mu}k_{1T}^{\nu\}}}{M_1} \big(\mathcal{H}^\perp_{1}+H^\perp_{1S}\big)\bar{\mathcal{H}}_{1T}
  -\frac{d_{T}^{\mu\nu}k_{1T}\cdot S_{2T}+k_{1T}^{\{\mu}S_{2T}^{\nu\}}}{M_1} \big(\mathcal{H}^\perp_{1}+{H}^\perp_{1S}\big)\bar{H}_{1T} \nonumber\\
  &-\frac{d_{T}^{\mu\nu}\varepsilon_{T k_2S_1}+\varepsilon_{T k_2}^{~~~\{\mu}S_{1T}^{\nu\}}}{M_2} H_{1T} \big(\bar{H}^\perp_{1}+\bar{\mathcal{H}}^\perp_{1S}\big)
  -\frac{d_{T}^{\mu\nu}k_{2T}\cdot S_{1T}+k_{2T}^{\{\mu}S_{1T}^{\nu\}}}{M_2} H_{1T} \big(\bar{\mathcal{H}}^\perp_{1}+\bar{H}^\perp_{1S}\big)\nonumber\\
  &-\frac{d_{T}^{\mu\nu}\varepsilon_{T S_1 k_2}+\varepsilon_{T S_1}^{~~~\{\mu}k_{2T}^{\nu\}}}{M_2} \mathcal{H}_{1T} \big(\bar{\mathcal{H}}^\perp_{1}+\bar{H}^\perp_{1S}\big)
  -\frac{d_{T}^{\mu\nu}k_{2T}\cdot S_{1T}+k_{2T}^{\{\mu}S_{1T}^{\nu\}}}{M_2} \mathcal{H}_{1T} \big(\bar{H}^\perp_{1}+\bar{\mathcal{H}}^\perp_{1S}\big)\nonumber\\
  &-\big(d_{T}^{\mu\nu}S_{1T}\cdot S_{2T}+S_{1T}^{\{\mu}S_{2T}^{\nu\}}\big)H_{1T}\bar H_{1T} -\big(d_{T}^{\mu\nu}\varepsilon_{T S_2S_1}-\varepsilon_{T S_2}^{~~~\{\mu}S_{1T}^{\nu\}}\big)H_{1T}\bar{\mathcal{H}}_{1T} \nonumber\\
  &-\big(d_{T}^{\mu\nu}S_{1T}\cdot S_{2T}+S_{1T}^{\{\mu}S_{2T}^{\nu\}}\big)\mathcal{H}_{1T}\bar{\mathcal{H}}_{1T} -\big(d_{T}^{\mu\nu}\varepsilon_{T S_2S_1}-\varepsilon_{T S_1}^{~~~\{\mu}S_{2T}^{\nu\}}\big)\mathcal{H}_{1T}\bar{H}_{1T}\bigg\} \Bigg\},\label{f:Trace}
\end{align}
\end{widetext}
where $d_T^{\mu\nu}=-g_T^{\mu\nu}$. 
Equation~(\ref{f:Trace}) is the complete hadronic tensor which is given in terms of both the $\mathcal{P}$ even and $\mathcal{P}$ odd fragmentation functions in SIA process. It can be easily checked that the hadronic tensor given in Eq.~(\ref{f:Trace}) satisfies the current coservation constraint, i.e., $q_\mu W^{\mu\nu}=q_\nu W^{\mu\nu}=0$.
Substituting Eq.~(\ref{f:Trace}) and Eq.~(\ref{f:Lmunu}) into Eq.~(\ref{f:cross-pp}), we can obtain the complete cross section in SIA process at leading twist.

\section{Cross sections and azimuthal asymmetries}\label{Sec-crosssection}

In the previous section, we show the complete hadronic tensor at leading twist in SIA process with both parity conserved FFs and parity violated FFs's contributions. In this section, we present the results by dividing the cross section into three parts according to polarizations of the produced dihadron, i.e., unpolarized part, single hadron polarized part and double hadron polarized part, respectively. The hadronic tensor shown in Eq.~(\ref{f:Trace}) is given in the transverse basis, to make Lorentz contraction with the leptonic tensor shown in Eq.~(\ref{f:Lmunu}), we should first transform it into the perpendicular bases. Here we will not present the calculation procedure and only show the final results.

\subsection{Unpolarized cross section and azimuthal asymmetries}

First of all we consider the unpolarized case where the hadron produced in final state are pseudoscalar particles or spin-0 particles. The differential cross section of it is given by
\begin{align}
   \frac{2E_1E_2d\sigma}{d^3 p_1d^3 p_2}&=\frac{8N_c\alpha^2_{em}e_q^2}{Q^4z_1z_2} \bigg[A(y) \Big( \mathcal{C}[D_1\bar D_1]-\mathcal{C}[\mathcal{D}_1 \bar{\mathcal{D}}_1]\Big) \nonumber\\
   &- B(y)\cos2\phi~ \mathcal{C}\Big[w_{12}\frac{H^\perp_1\bar H^\perp_1+\mathcal{H}^\perp_1\bar{\mathcal{H}}^\perp_1}{M_1M_2}\Big]\nonumber\\
   &- B(y)\sin2\phi~ \mathcal{C}\Big[w_{12}\frac{H^\perp_1\bar{\mathcal{H}}^\perp_1-\mathcal{H}^\perp_1\bar H^\perp_1}{M_1M_2}\Big]\bigg], \label{f:crossUn}
\end{align}
where the abbreviation $\mathcal{C}[D_1\bar D_1]$ denotes a convolution,
\begin{align}
  \mathcal{C}[D_1\bar D_1]=\int \frac{d^2 k_{1T}}{(2\pi)^2}\frac{d^2 k_{2T}}{(2\pi)^2}\delta^2( \vec q_T- \vec k_{1T}- \vec k_{2T}) D_1\bar D_1. \label{f:convolution}
\end{align}
The wight function $w_{12}=2\hat{\mathbf{h}}\cdot \mathbf{k}_{1T}\hat{\mathbf{h}}\cdot \mathbf{k}_{2T}-\mathbf{k}_{1T}\cdot \mathbf{k}_{2T}/M_1M_2$.
From Eq.~(\ref{f:crossUn}) we can see that the $\cos2\phi$ terms corresponds to the Collins effect which is the reflection of the Collins function $H^\perp_1$ ($\bar H^\perp_1$). Apart from the contribution of the Collins function, there is also contribution of $\mathcal{P}$ odd Collins-type function $\mathcal{H}_1^\perp$ ($\bar{\mathcal{H}}_1^\perp$). This azimuthal asymmetry can be written as
\begin{align}
  A_{UU}^{\cos2\phi}=-\frac{B(y)}{A(y)}\frac{\mathcal{C}\big[w_{12}\big(H^\perp_1\bar H^\perp_1+\mathcal{H}^\perp_1\bar{\mathcal{H}}^\perp_1\big)\big]}{\mathcal{C}[D_1\bar D_1]-\mathcal{C}[\mathcal{D}_1 \bar{\mathcal{D}}_1]}, \label{f:CollinsA}
\end{align}
where the subscript $UU$ denotes unpolarized cases. The first $U$ denotes hadron 1 is unpolarized, the same to the second hadron.
Often we define the weighted functions, i.e.,
\begin{align}
  &D_1(z)=\int \frac{d^2k_T}{(2\pi)^2}D_1(z, k_T), \\
  &H_1^\perp(z)=\int \frac{d^2k_T}{(2\pi)^2}\frac{|k_T|}{M}H_1^\perp(z, k_T).
\end{align}
Then we have
\begin{align}
  \tilde A_{UU}^{\cos2\phi}=-\frac{B(y)}{A(y)}\frac{H^\perp_1\bar H^\perp_1+\mathcal{H}^\perp_1\bar{\mathcal{H}}^\perp_1}{D_1\bar D_1-\mathcal{D}_1 \bar{\mathcal{D}}_1}. \label{f:CollinsAWeight}
\end{align}
We notice that the sign before two term in the numerator are the same, which means both the Collins function and Collins-type function have positive contributions to the $\cos2\phi$ azimuthal asymmetry. In other words Collins effect ($\cos2\phi$) may have alternative origins except for the Collins function as indicated in ref.~\cite{Kang:2010qx}. The $\mathcal{P}$ odd FFs will complete the extraction of Collins function, in other words the Collins function extracted from experiments data contains $\mathcal{H}^\perp_1$ which is not easily to excluded.

Apart from the $\cos2\phi$ azimuthal asymmetry, we also obtain one special azimuthal asymmetry at leading twist, i.e., $\sin2\phi$ asymmetry. $\sin2\phi$ azimuthal asymmetry arise from the interference of $H^\perp_1$ ($\bar H^\perp_1$) and $\bar{\mathcal{H}}^\perp_1$ ($\mathcal{H}^\perp_1$), i.e.,
\begin{align}
 \tilde  A_{UU}^{\sin2\phi}=-\frac{B(y)}{A(y)}\frac{H^\perp_1\bar{\mathcal{H}}^\perp_1-\mathcal{H}^\perp_1\bar H^\perp_1}{D_1\bar D_1-\mathcal{D}_1 \bar{\mathcal{D}}_1}. \label{f:CollinsInt}
\end{align}
$\tilde A_{UU}^{\sin2\phi}$ vanishes if only $\mathcal{P}$-even FFs are taken into consideration as discussed in ref.~\cite{Chen:2016moq}. The first term in the numerator corresponds to the interference between Collins function $H_1^\perp$ and
$\mathcal{P}$-odd (anti) Collins-type  FF $\bar{\mathcal{H}}^\perp_1$ fragmented by antiquark while the second terms corresponds to the interference between (anti) Collins function $\bar H_1^\perp$ and
$\mathcal{P}$ odd Collins-type FF $\mathcal{H}^\perp_1$ fragmented by quark. We also notice that the signs before these two terms in the numerator are opposite, which means that the effect will vanish when sum over many events. In other words we conclude that the $\mathcal{P}$ odd effect only survive on the event-by-event basis.

\subsection{Single hadron polarized cross section and asymmetries}

For the single hadron polarized case, we assume that hadron 1 is polarized while hadron 2 is unpolarized, or hadron 1 is spin-1/2 particle while hadron 2 is spin-0 particle. For simplicity, we omit those term which have been shown in Eq.~(\ref{f:crossUn}). Thus the cross section is given by
\begin{align}
  \frac{2E_1E_2d\sigma}{d^3 p_1d^3 p_2}&=\frac{8N_c\alpha^2_{em}e_q^2}{Q^4z_1z_2}\Bigg\{A(y) \Big( \mathcal{C}[D_1\bar D_1]-\mathcal{C}[\mathcal{D}_1 \bar{\mathcal{D}}_1]\Big)\nonumber\\
  +A(y)&\bigg\{\lambda_1\Big[\mathcal{C}[\mathcal{G}_{1L}\bar D_1]-\mathcal{C}[{G}_{1L}\bar {\mathcal D}_1]\Big] \nonumber\\
  +&|S_{1T}| \Big\{\cos(\phi-\phi_{S_1})\mathcal{C}\big[ w_1 \big(\mathcal{D}^\perp_{1T}\bar D_1-G^\perp_{1T}\bar{\mathcal{D}}_1\big)\big] \nonumber\\
  &\hspace{5mm}+\sin(\phi-\phi_{S_1})\mathcal{C}\big[ w_1 \big(\mathcal{G}^\perp_{1T}\bar{ \mathcal{D}}_1-D^\perp_{1T}\bar D_1\big)\big]\Big\}\bigg\} \nonumber\\
  -B(y)&\bigg\{\lambda_1 \sin2\phi \big[w_{12}\big(H^\perp_{1L}\bar H_1^\perp-\mathcal{H}^\perp_{1L}\mathcal{H}^\perp_1\big)\big]\nonumber\\
  &+\lambda_1 \cos2\phi \big[w_{12}\big(\mathcal{H}^\perp_{1L}\bar H_1^\perp+H^\perp_{1L}\mathcal{H}^\perp_1\big)\big]\nonumber\\
  +&|S_{1T}| \Big\{\cos(\phi+\phi_{S_1})\mathcal{C}\big[ w_2 \big(H_{1T}\bar {\mathcal{H}}^\perp_1+\mathcal{H}_{1T}\bar H_1^\perp\big)\big] \nonumber\\
  &\hspace{5mm}+\sin(\phi+\phi_{S_1})\mathcal{C}\big[ w_2 \big(H_{1T}\bar H_1^\perp-\mathcal{H}_{1T}\bar {\mathcal{H}}^\perp_1\big)\big]\nonumber\\
  +\cos2\phi&\cos(\phi-\phi_{S_1})\mathcal{C}\big[w_{12} w_1 \big(\mathcal{H}^\perp_{1T}\bar H_1^\perp+H^\perp_{1T}\bar {\mathcal{H}}^\perp_1\big)\big]\nonumber\\
  +\sin2\phi&\cos(\phi-\phi_{S_1})\mathcal{C}\big[w_{12} w_1 \big(\mathcal{H}^\perp_{1T}\bar {\mathcal{H}}^\perp_1-H^\perp_{1T}\bar H_1^\perp\big)\big]\Big\}\bigg\}
  \Bigg\}, \label{f:singlecross}
\end{align}
where the weighted functions $w_1=\hat{\mathbf{h}}\cdot \mathbf{k}_{1T}/M_1$, $w_2=\hat{\mathbf{h}}\cdot \mathbf{k}_{2T}/M_2$. From Eq.~(\ref{f:singlecross}) we can see that the transverse polarization part vanish when integrate over the transverse momenta. However, we here need to the consider azimuthal asymmetries and the transverse momenta should not be integrated over except for weighted integrals. Here we have,
\begin{align}
 & A_{TU}^{\sin(\phi-\phi_{S_1})}=\frac{\mathcal{C}\big[ w_1 \big(\mathcal{G}^\perp_{1T}\bar{\mathcal{D}}_1-D^\perp_{1T}\bar D_1\big)\big]}{\mathcal{C}[D_1\bar D_1]-\mathcal{C}[\mathcal{D}_1 \bar{\mathcal{D}}_1]}, \label{f:sinphi-phis1}  \\
 & A_{TU}^{\cos(\phi-\phi_{S_1})}=\frac{\mathcal{C}\big[ w_1 \big(\mathcal{D}^\perp_{1T}\bar D_1-G^\perp_{1T}\bar{\mathcal{D}}_1\big)\big]}{\mathcal{C}[D_1\bar D_1]-\mathcal{C}[\mathcal{D}_1 \bar{\mathcal{D}}_1]}. \label{f:cosphi-phis1}
\end{align}
From Eq.~(\ref{f:sinphi-phis1}) we can see that $A_{UL}^{\sin(\phi-\phi_{S_1})}$ corresponds to the Sivers effect in fragmentation process since this azimuthal asymmetry comes from the FFs $D^\perp_{1T}\bar D_1$. But we see the two terms in the numerator have opposite signs which means that the $\mathcal{P}$ odd FFs $\mathcal{G}^\perp_{1T}\bar{\mathcal{D}}_1$ have negative contribution to this asymmetry relative to $D^\perp_{1T}\bar D_1$. Thus the $\mathcal{P}$ odd FFs can reduce the magnitude of the asymmetry. However, azimuthal asymmetry $A_{UL}^{\cos(\phi-\phi_{S_1})}$ comes from the interference of the $\mathcal{P}$-odd FF(s) $\mathcal{D}^\perp_{1T} (\bar{\mathcal{D}}_1 )$ and $\mathcal{P}$-even FF(s) $\bar D_1 (G^\perp_{1T})$ and it only survive event-by-event basis as we have discussed for Eq.~(\ref{f:CollinsAWeight}).

Besides, we can also calculate the other two azimuthal asymmetries which are generated from the chiral-odd FFs. (The following two azimuthal asymmetries are not the right simplified ones, we show them here for direct impression.) They are given by,
\begin{align}
 & A_{TU}^{\sin(\phi+\phi_{S_1})}=-\frac{\mathcal{C}\big[ w_2 \big(H_{1T}\bar H_1^\perp-\mathcal{H}_{1T}\bar {\mathcal{H}}^\perp_1\big)\big]}{\mathcal{C}[D_1\bar D_1]-\mathcal{C}[\mathcal{D}_1 \bar{\mathcal{D}}_1]}, \label{f:sinphi+phis1}  \\
 & A_{TU}^{\cos(\phi+\phi_{S_1})}=-\frac{\mathcal{C}\big[ w_2 \big(H_{1T}\bar {\mathcal{H}}^\perp_1+\mathcal{H}_{1T}\bar H_1^\perp\big)\big]}{\mathcal{C}[D_1\bar D_1]-\mathcal{C}[\mathcal{D}_1 \bar{\mathcal{D}}_1]}. \label{f:cosphi+phis1}
\end{align}
We can see these two asymmetries are related to the transverse polarization transformation function $H_{1T}$ and Collins function $H_1^\perp$. Similar to $A_{LU}^{\sin(\phi-\phi_{S_1})}$, the contribution of $\mathcal{P}$ odd FFs $\mathcal{H}_{1T}\bar {\mathcal{H}}^\perp_1$ can reduce the contribution from $\mathcal{P}$ even FFs $H_{1T}\bar H_1^\perp$, since they have different signs.
It is interesting to see that the interference two terms contributing to azimuthal asymmetry $A_{LU}^{\cos(\phi+\phi_{S_1})}$ have both positive contributions and they can survive even summing over many events. At leading twist, $ A_{TU}^{\cos(\phi+\phi_{S_1})}$ vanishes for parity violated reason, see ref~\cite{Chen:2016moq}. Thus $ A_{TU}^{\cos(\phi+\phi_{S_1})}$ is important for determining the $\mathcal{P}$-odd FFs by measuring the azimuthal asymmetry at leading twist.

\subsection{Double hadron polarized cross section and asymmetries}

For the double hadron polarized case, we consider two spin-1/2 hadrons in the final states, there are 54 terms. For simplicity, we only consider the integrated cross section and only a few of them are left which are given by
\begin{align}
  &\frac{d\sigma}{dz_1dz_2dy}=\frac{4\pi N_ce_q^2\alpha_{em}^2}{Q^2}\Big[A(y) \big( D_1\bar D_1-\mathcal{D}_1 \bar{\mathcal{D}}_1\big) \nonumber\\
  &\hspace{25mm}-A(y)\lambda_1\lambda_2\big(G_{1L}\bar G_{1L}-\mathcal{G}_{1L}\bar{\mathcal{G}}_{1L}\big)\nonumber\\
  &-B(y)|S_{1T}||S_{2T}|\cos(\phi_{S_1}+\phi_{S_2})\big(H_{1T}\bar H_{1T}+\mathcal{H}_{1T}\bar{\mathcal{H}}_{1T}\big)\nonumber\\
  &-B(y)|S_{1T}||S_{2T}|\sin(\phi_{S_1}+\phi_{S_2})\big(H_{1T}\bar{\mathcal{H}}_{1T}-\mathcal{H}_{1T}\bar H_{1T}\big)\Big]. \label{f:cross-ss}
\end{align}
Here we have used
\begin{align}
  d^3p_1d^3p_2/E_1E_2&=(dz_1/z_1)(z_2Q^2dz_2/4)d\mathbf{p}_{1\perp}d\Omega_2 \nonumber\\
  &=\pi Q^2z_1z_2dz_1dz_2dyd^2\mathbf{q}_T.
\end{align}

In Eq.~(\ref{f:cross-ss}), there are two azimuthal asymmetries which correspond to the double spin asymmetries.
\begin{align}
  & \tilde A_{TT}^{\cos(\phi_{S_1}+\phi_{S_2})} =-\frac{B(y)}{A(y)}\frac{H_{1T}\bar H_{1T}+\mathcal{H}_{1T}\bar{\mathcal{H}}_{1T}}{D_1\bar D_1-\mathcal{D}_1 \bar{\mathcal{D}}_1}, \label{f:DoubelspinCos}\\
  & \tilde A_{TT}^{\sin(\phi_{S_1}+\phi_{S_2})} =-\frac{B(y)}{A(y)}\frac{H_{1T}\bar{\mathcal{H}}_{1T}-\mathcal{H}_{1T}\bar H_{1T}}{D_1\bar D_1-\mathcal{D}_1 \bar{\mathcal{D}}_1}. \label{f:DoubelspinSin}
\end{align}
First of all we consider the asymmetry $\tilde A_{TT}^{\cos(\phi_{S_1}+\phi_{S_2})}$. We find that both $H_{1T}\bar H_{1T}$ and $\mathcal{H}_{1T}\bar{\mathcal{H}}_{1T}$ have contributions to this asymmetry which means $\tilde A_{TT}^{\cos(\phi_{S_1}+\phi_{S_2})}$ has one more origin and it will complicate the extraction of the transverse polarization transformation  function $H_{1T}$. This is the same to Collins asymmetry, $\tilde A_{UU}^{\cos2\phi}$.  The asymmetry $\tilde A_{TT}^{\sin(\phi_{S_1}+\phi_{S_2})} $, similar to $\tilde A_{UU}^{\sin2\phi}$, arise from the two interference terms with opposite signs. So it can only survive on event-by-event basis.

So far we present eight kinds of azimuthal asymmetries. We can divide them into four parts according to the FFs' contributions.
\begin{itemize}
  \item $\tilde A_{UU}^{\cos2\phi}$ and $\tilde A_{TT}^{\cos(\phi_{S_1}+\phi_{S_2})}$. In this part, both the $\mathcal{P}$ odd FFs and $\mathcal{P}$ even FFs have positive contributions which means these kinds of azimuthal asymmetries have two origins.

  \item $\tilde A_{UU}^{\sin2\phi}$, $\tilde A_{TU}^{\cos(\phi-\phi_{S_1})}$ and $\tilde A_{TT}^{\sin(\phi_{S_1}+\phi_{S_2})} $. In this part, asymmetries comes from the interference between the $\mathcal{P}$ odd FF and $\mathcal{P}$ even FF and terms in the numerator have opposite signs which means these azimuthal asymmetries only survive event-by-event.

  \item $\tilde A_{TU}^{\sin(\phi-\phi_{S_1})}$ and $A_{TU}^{\sin(\phi+\phi_{S_1})}$. These asymmetries are from both the $\mathcal{P}$ odd FFs and $\mathcal{P}$ even FFs but these two terms have opposite sign. In other words, the $\mathcal{P}$ odd FFs can reduce the effect generated by $\mathcal{P}$ even FFs.

  \item $A_{TU}^{\cos(\phi+\phi_{S_1})}$. This kind of azimuthal asymmetry is very special. The interference two terms contributing to azimuthal asymmetry $A_{LU}^{\cos(\phi+\phi_{S_1})}$ have both positive contributions and they can survive even summing over many events. This is important for determining the $\mathcal{P}$-odd FFs by measuring the azimuthal asymmetry at leading twist.
\end{itemize}

\subsection{Hadron polarizations}

In the previous sections we considered azimuthal asymmetries which can be detected directly in SIA experiments. If the spin of hadrons measured in the final state is not zero, the hadron polarization can be measured. For simplicity, in this part we only consider the one hadron polarized case. The hadron polarizations can be calculated in three steps. The first one is to illustrate the probability interpretation, the second is to calculate eigenvalues and the third is the to calculate hadron polarizations. In the following we will take $\langle S_T^x\rangle$ for example to illustrate the calculations.

Hadron polarization $\langle S_T^x\rangle$, in general, denotes the probability of the hadrons in the spin $S_T^x$ state. For a system of hadrons, we use spin density matrix to describe the polarization. The spin density matrix can be written as
\begin{align}
  \rho=\sum_{m,n}|n\rangle \langle n|~ \rho ~|m\rangle \langle m|=\sum_{m,n}\mathrm{Tr}\big[\rho |m\rangle \langle n|\big]|n\rangle \langle m|,\label{f:spindensity}
\end{align}
where $|m\rangle, |n\rangle$ are eigenstates of corresponding spin operators and $P\equiv\mathrm{Tr}\big[\rho |m\rangle \langle n|\big]$ is the probability of finding one of these states.  The spin operator can be defined in arbitrary direction, i.e.,
\begin{align}
  \Sigma^i\hat n^i=\Sigma^x \sin\theta\cos\phi +\Sigma^y \sin\theta\sin\phi +\Sigma^z \cos\theta, \label{f:spinoperator}
\end{align}
where $\theta$ and $\phi$ denote the polar and azimuthal angle. Thus the polarization of a system can be calculated as
\begin{align}
  O=\langle \mathcal{O} \rangle =\mathrm{Tr} [\mathcal{O}\rho], \label{f:spinO}
\end{align}
where $\mathcal{O}$ denotes any spin operator. Using Eq.~(\ref{f:spindensity})-(\ref{f:spinO}), we obtain
\begin{align}
  & \langle S_T^x \rangle=P(1,\frac{\pi}{2},0)-P(-1,\frac{\pi}{2},0), \label{f:stxp}
\end{align}
where $P=P(m,\theta,\phi)$ where $m$ denotes the eigenvalue of the corresponding state. Do not be confused with the eigenstate $|m \rangle$. So $ \langle S_T^x \rangle $ denote the difference of the two probabilities. We can also calculate
\begin{align}
  & \langle \lambda \rangle=P(1,0,0)-P(-1,0,0), \label{f:lambdap}\\
  & \langle S_T^y \rangle=P(1,\frac{\pi}{2},\frac{\pi}{2})-P(-1,\frac{\pi}{2},\frac{\pi}{2}). \label{f:stxp}
\end{align}
 Similarly, all the eigenvalues of the system on eigenstates of $\lambda, S_T^x, S_T^y$ can be calculated. They are given in Table.~\ref{table1}.

\begin{table}[ht]
 \renewcommand\arraystretch{1.5}
\begin{tabular}{|c||c|c|c|c|c|c|}
  \hline
             &~ $|\psi_{z+}\rangle$~~ &~ $ |\psi_{z-} \rangle $ ~~& ~$|\psi_{y+}\rangle$ ~~& ~~$ |\psi_{y-} \rangle $ ~~& ~~$|\psi_{x+}\rangle$ ~~& ~~$ |\psi_{x-} \rangle $ ~~  \\  \hline \hline
  $ \lambda$ & $1$ & $-1$ & $0$ & $0$  & $0$ & $0$   \\ \hline
  $ ~~S_T^x ~~ $ & $0$ & $0$  & $1$ & $-1$ & $0$ & $0$  \\ \hline
  $ S_T^y  $ & $0$ & $0$  & $0$ & $0$  & $1$ & $-1$  \\ \hline
\end{tabular} \caption{Eigenvalues of the system on eigenstates of $\lambda, S_T^x, S_T^y$.} \label{table1}
\end{table}

To calculate the hadron polarization, we consider the single hadron polarized case. First of all, we consider the longitudinal polarization. Using Eq.~(\ref{f:lambdap}) and substituting the corresponding eigenvalues in Table.~\ref{table1} into Eq.~(\ref{f:singlecross}), we can obtain the longitudinal polarization of hadron in SIA process,
\begin{align}
  \langle \lambda_1 \rangle =\frac{\mathcal{C}[\mathcal{G}_{1L}\bar D_1]-\mathcal{C}[{G}_{1L}\bar {\mathcal D}_1]}{\mathcal{C}[D_1\bar D_1]-\mathcal{C}[\mathcal{D}_1 \bar{\mathcal{D}}_1]}.\label{f:longitudinalP}
\end{align}
From Eq.~(\ref{f:longitudinalP})we see the longitudinal polarization in SIA process arising form the interference of the $\mathcal{P}$ odd FF(s) $\mathcal{G}_{1L}$ ($\bar {\mathcal D}_1$) and the $\mathcal{P}$ even FF(s) $\bar D_1$ (${G}_{1L}$). The signs of the terms in the numerator are opposite which means that the $\langle \lambda_1 \rangle$ only survives on event-bey-event basis because the averaged over cross section of many event vanishes.

According to the cross section shown in Eq.~(\ref{f:singlecross}) we know the transverse polarization cannot be calculate directly. In Fig.~\ref{Frame}, we define the plane determined by momenta $l_1, l_2$ and $p_2$ as $x-O-z$ plane or lepton-hadron plane. $p_2$ lies in z-direction while the x-direction is determined by the transverse momenta of leptons as shown. Here we say transverse polarization is the polarization defined with respect to the lepton hadron plane. Chen $et~al.$ argued in ref.~\cite{Chen:2016moq,Yang:2017sxz}, the transverse hadron polarization $\langle S_T^x \rangle$ corresponds to $|S_{T}|\cos\phi_{S}$ while $\langle S_T^y \rangle$ corresponds to $|S_{T}|\sin\phi_{S}$ and these polarizations vanish in Eq.~(\ref{f:singlecross}). In fact, the transverse polarization vanish at leading twist but they survive at twist-3 level. However, in ref.~\cite{Chen:2016moq,Yang:2017sxz}, Chen $et~al.$ also discussed the transverse polarizations with respect to hadron-hadron place which is defined by the momenta of the two hadrons as shown in Fig.~\ref{Frame}. In the case we can define unit vector $\vec e_n$ and $\vec e_t$ to label the transverse directions. Here $\vec e_n=\vec p_1\times \vec p_2/|\vec p_1\times \vec p_2|=(-\sin\phi, \cos\phi)$ and $\vec e_t= \vec p_{1T}/|\vec p_{1T}|=(\cos\phi, \sin\phi)$, i.e., the normal and tangent of the hadron-hadron plane, respectively. Thus the transverse hadron polarization $S_T^n$ corresponds to $|S_{T}|\sin(\phi_{S}-\phi)$ while $S_T^t$ corresponds to $|S_{T}|\cos(\phi_{S}-\phi)$. So we can calculate the transverse polarizations with respect to hadron hadron plane.
\begin{align}
  \langle S_T^n \rangle =-\frac{\mathcal{C}\Big[ w_1 \big(\mathcal{G}^\perp_{1T}\bar{\mathcal{D}}_1-D^\perp_{1T}\bar D_1 \big)\Big]}{\mathcal{C}[D_1\bar D_1]-\mathcal{C}[\mathcal{D}_1 \bar{\mathcal{D}}_1]}, \label{f:stn} \\
  \langle S_T^t \rangle =\frac{\mathcal{C}\Big[ w_1 \big(\mathcal{D}^\perp_{1T}\bar D_1-G^\perp_{1T}\bar{\mathcal{D}}_1 \big)\Big]}{\mathcal{C}[D_1\bar D_1]-\mathcal{C}[\mathcal{D}_1 \bar{\mathcal{D}}_1]}. \label{f:stt}
\end{align}
For $\langle S_T^n \rangle$, we see it reflects the effect of Sivers-type FF $D_{1T}^\perp$. But the two terms in the numerator have opposite signs, which means that the contribution from $\mathcal{P}$-odd FFs $\mathcal{G}^\perp_{1T}\bar{\mathcal{D}}_1$ will dilute the contribution from $\mathcal{P}$-even FFs $D^\perp_{1T}\bar D_1$. However the transverse polarization $\langle S_T^t \rangle$ comes from the interference of the $\mathcal{P}$-odd FF(s) $\mathcal{D}^\perp_{1T} (\bar{\mathcal{D}}_1 )$ and $\mathcal{P}$-even FF(s) $\bar D_1 (G^\perp_{1T})$ and it only survive event-by-event basis, just like $\langle \lambda_1 \rangle$, see ref~\cite{Chen:2016moq}.

\section{Summary}\label{Sec-summary}

Fragmentation functions are important quantities in high energy reactions and they describe the fragmentation and/or hadronization processes. FFs can be defined by the quark-quark correlation functions which satisfy the hermiticity and parity conservation constraints, in general. When the non-trivial QCD topological structure is taken into consideration, QCD theory can be local parity violated. In this case, the parity conservation constraint will break down. As a result, FFs can be $\mathcal{P}$-odd in addition to the $\mathcal{P}$-even properties. We apply this argument into the SIA process in order to calculate the azimuthal asymmetries and hadron polarizations which are expressed in terms of FFs including both the $\mathcal{P}$-odd ones and the $\mathcal{P}$-even ones.

We calculate eight kinds of azimuthal asymmetries and category them into four parts.
The first part is $\tilde A_{UU}^{\cos2\phi}$ and $\tilde A_{TT}^{\cos(\phi_{S_1}+\phi_{S_2})}$. In this part, both the $\mathcal{P}$-odd FFs and $\mathcal{P}$-even FFs have positive contributions which means these kinds of azimuthal asymmetries have two origins.
The second part is $\tilde A_{UU}^{\sin2\phi}$, $\tilde A_{TU}^{\cos(\phi-\phi_{S_1})}$ and $\tilde A_{TT}^{\sin(\phi_{S_1}+\phi_{S_2})} $. In this part, asymmetries comes from the interference between the $\mathcal{P}$-odd FF and $\mathcal{P}$-even FF and terms in the numerator have opposite signs which means these azimuthal asymmetries only survive event-by-event.
The third part is $\tilde A_{TU}^{\sin(\phi-\phi_{S_1})}$ and $A_{TU}^{\sin(\phi+\phi_{S_1})}$. These asymmetries are from both the $\mathcal{P}$-odd FFs and $\mathcal{P}$-even FFs but these two terms have opposite sign. In other words, the $\mathcal{P}$-odd FFs can reduce the effect generated by $\mathcal{P}$-even FFs.
The fourth part is $A_{TU}^{\cos(\phi+\phi_{S_1})}$. This kind of azimuthal asymmetry is very special. The interference two terms contributing to azimuthal asymmetry $A_{LU}^{\cos(\phi+\phi_{S_1})}$ have both positive contributions and they can survive even summing over many events. This is important for determining the $\mathcal{P}$-odd FFs by measuring the azimuthal asymmetry at leading twist.

We also calculate the hadron polarizations in which we assume that hadron 1 is polarized and hadron 2 is unpolarized. We find that the longitudinal polarization in SIA process arising form the interference of the $\mathcal{P}$-odd FF(s) $\mathcal{G}_{1L}$ ($\bar {\mathcal D}_1$) and the $\mathcal{P}$-even FF(s) $\bar D_1$ (${G}_{1L}$). The signs of the terms in the numerator are opposite which means that the $\langle \lambda_1 \rangle$ only survives on event-bey-event basis because the averaged over cross section of many event vanishes.
For $\langle S_T^n \rangle$, we see it reflects the effect of Sivers-type FF $D_{1T}^\perp$. But the two terms in the numerator have opposite signs, which means that the contribution from $\mathcal{P}$-odd FFs $\mathcal{G}^\perp_{1T}\bar{\mathcal{D}}_1$ will dilute the contribution from $\mathcal{P}$-even FFs $D^\perp_{1T}\bar D_1$. However the transverse polarization $\langle S_T^t \rangle$ comes from the interference of the $\mathcal{P}$-odd FF(s) $\mathcal{D}^\perp_{1T} (\bar{\mathcal{D}}_1 )$ and $\mathcal{P}$-even FF(s) $\bar D_1 (G^\perp_{1T})$ and it only survive event-by-event basis.

\end{document}